%% file: JAI2015-intro.tex
\documentclass{ws-jai}
\usepackage[flushleft]{threeparttable}
\usepackage{natbib}
\usepackage{color}

\newcommand{\rom}[1]{\uppercase\expandafter{\romannumeral #1\relax}}

\begin{document}

\catchline{}{}{}{}{} 

\markboth{Christe S. et al.}{Sounding Rockets}

\title{Introduction to the Special Issue on Sounding Rockets and Instrumentation}

\author{Steven Christe$^{1,\dagger}$, 
	Ben Zeiger$^2$,
   	Rob Pfaff$^1$,
	Michael Garcia$^3$
}

\address{
$^1$NASA Goddard Space Flight Center, Greenbelt, MD 20771, USA, steven.christe@nasa.gov\\
$^2$University of Colorado, CO 80303, USA\\
$^3$NASA Headquarters, Washington, DC 20546, USA \\
}
\maketitle

\corres{$^\dagger$Corresponding author.}

\begin{history}
\received{(to be inserted by publisher)};
\revised{(to be inserted by publisher)};
\accepted{(to be inserted by publisher)};
\end{history}

\begin{abstract}
Rocket technology, originally developed for military applications, has provided a low-cost observing platform to carry critical and rapid-response scientific investigations for over 70 years. Even with the development of launch vehicles that could put satellites into orbit, high altitude sounding rockets have remained relevant. In addition to science observations, sounding rockets provide a unique technology test platform and a valuable training ground for scientists and engineers. Most importantly, sounding rockets remain the only way to explore the tenuous regions of the Earth's atmosphere (the upper stratosphere, mesosphere, and lower ionosphere/thermosphere) above balloon altitudes ($\sim$40 km) and below satellite orbits ($\sim$160 km). They can lift remote sensing telescope payloads with masses up to 400 kg to altitudes of 350 km providing observing times of up to 6 minutes above the blocking influence of Earth's atmosphere.
Though a number of sounding rocket research programs exist around the world, this article focuses on the NASA Sounding Rocket Program, and particularly on the astrophysical and solar sounding rocket payloads.
\end{abstract}


\section{Introduction}
\label{sec:introduction}
\input{sec-introduction}

\section{The History of Sounding Rockets}
\label{sec:history}
\input{sec-history}

\section{Sounding Rocket Capabilities for Solar and Astrophysical Payloads}
\label{sec:capabilities}
The following sections provide an overview of the capabilities of modern sounding rockets supported by NASA's Sounding Rocket Program most relevant to solar and astrophysics payloads. The SRPO user's manual \citep{NSROC:2015va} provides much more detail than is provided here. An article in the 17th ESA Symposium on European Rocket and Balloon Programmes and Related Research conference proceedings \citep{2005ESASP.590..305K} also provides a good overview of the program.
\subsection{Typical Rockets, Payload Configurations, and Timelines }
\label{sec-capabilities-rockets}
\input{sec-capabilities-rockets}

\subsection{Launch Sites}
\label{sec-capabilities-launch-sites}
\input{sec-capabilities-launch-sites}
\subsection{Pointing Systems}
The NASA Sounding Rocket team at WFF provides experimenters with pre-built and high heritage pointing systems. For solar-pointing payloads the SPARCS system is provided while for astrophysics payloads it is the Celestial ACS. Both of these systems handle all aspects of pointing including control and pointing knowledge. The primary pointing sensors for both systems are generally integrated in the experiment section and aligned to the experimenters telescope. 

\subsubsection{Solar Pointing Attitude Rocket Control System (SPARCS VII)}
\input{sec-capabilities-pointing-sparcs}

\subsubsection{Celestial Attitude Control System (CACS)}
\input{sec-capabilities-pointing-star}

\subsection{Telemetry \& Commanding}
\label{sec-capabilities-telemetry}
\input{sec-capabilities-telemetry}

\subsection{Recovery}
\input{sec-capabilities-recovery}

\section{Sounding Rockets as a Science Platform}
\label{sec:science-platform}
\input{sec-science-platform}

\section{Examples of Solar and Astrophysics Science Payloads}
\label{sec:science}
In the sections that follow, we provide a few examples of recent sounding rocket experiments. Additional articles in this issue provide in-depth studies of several sounding rocket payloads.
\subsection{Solar}
\input{sec-solar-payloads}
\label{sec:solar-science}

\subsection{Astrophysics}
\label{sec:astro-science}

\input{sec-astro-payloads}

\section{Future Developments \& Conclusions}
\label{sec:future}
\input{sec-future}

\section*{Acknowledgments}
We would like to acknowledge the help of Philip Eberspeaker the Chief of the NASA Sounding Rockets Program Office for providing
the launch database. We would also like to acknowledge thank the NASA SRPO and all of the employees
of the NASA Sounding Rocket Operations Contract (NSROC) Program who work hard to support the NASA sounding rocket program. Additionally, this research made use of NASA's Astrophysics Data System. Finally,
we'd like to thank all of the scientists and engineers who use this platform and have made it a success.
%
%
\bibliographystyle{ws-jai}
\bibliography{bib-sdc-references,bib-brz-references}
\end{document}

%% file: sec-introduction.tex



Rocket technology pushed upward quickly in the first half of the 20th Century, from achieving altitudes of tens of meters to a few kilometers in the 1920s to hundreds of kilometers in the 1940s. Even with the development of launch vehicles that could put satellites into orbit, however, high altitude sounding rockets remain relevant. They provide the relatively inexpensive proving ground between terrestrial observations, instrumentation and orbiting observatories. Despite the short duration of flights compared to orbital observatories, sounding rockets are low-cost vehicles offering critical science observations, rapid-response missions, unique technology test platforms, and valuable training grounds for scientists and engineers.

Although this overview article focuses on sounding rocket use for solar physics and astrophysics research, the continued scientific impact of sounding rockets covers a broad spectrum of disciplines, also including Earth science, geophysics, space physics, planetary science, and microgravity research.  Most importantly, sounding rockets remain the only way to explore the tenuous regions of the Earth's atmosphere (the upper stratosphere, mesosphere, and lower ionosphere/thermosphere) above balloon altitudes ($\sim$40 km) and below satellite orbits ($\sim$160 km). 

Though a number of sounding rocket research programs exist around the world, this article focuses on the NASA Sounding Rocket Program, and particularly on the astrophysical and solar sounding rocket payloads. The program is managed and executed at the Wallops Flight Facility (WFF) in Virginia (USA), which is part of the NASA/Goddard Space Flight Center.

%% file: sec-history.tex

The following introduction, unless otherwise cited, is largely based on the introductions to rocket research of \citet{1971NASSP4401.....C} and \citet{ordway2008rocket}.

The history of rocketry is a long and storied one, driven by military needs for centuries before becoming of scientific value. The first rockets were likely developed by the Chinese for military purposes and were a natural progression from their invention of gunpowder to propel objects such as cannon balls and arrows. The first use of such ``fire arrows'' appears as early as 904-907 A.D. in the {\it Records of the Nine Kingdoms}, documenting the siege of Yuzhang, China.
Perhaps the first time rockets were recorded to have been used in mainland Europe was during the Battle of Chioggia between the Genoese and Venetians in 1380 \citep{chiogga}. This was also when the modern name for rockets was first published from the Italian {\it rocchetta}, which means ``little fuse'' \citep{1975hrst.book.....V}.

\begin{figure}[ht]
\centering
\begin{minipage}[b]{0.45\linewidth}
  \includegraphics[width=3in]{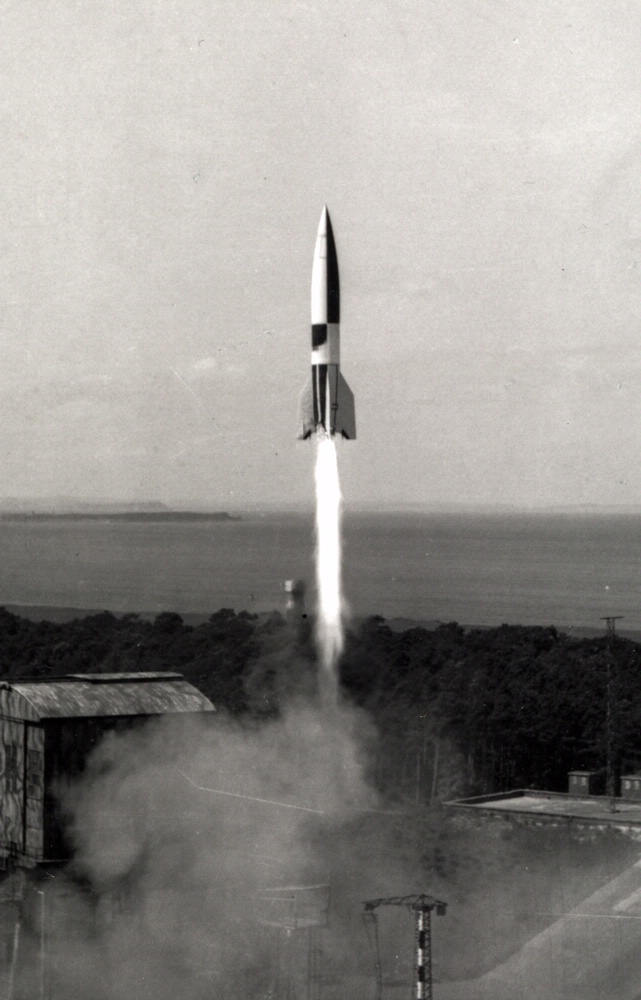}
  \caption{The launch of a German A4 rocket, later to be renamed V-2, at Peenm\"{u}nde, Germany, 1940s. Image courtesy of the German Federal Archives (Bundesarchiv, Bild 141-1880 / CC-BY-SA 3.0).}
  \label{fig:v2-launch-photo}
\end{minipage}
\quad
\begin{minipage}[b]{0.45\linewidth}
  \includegraphics[width=3in]{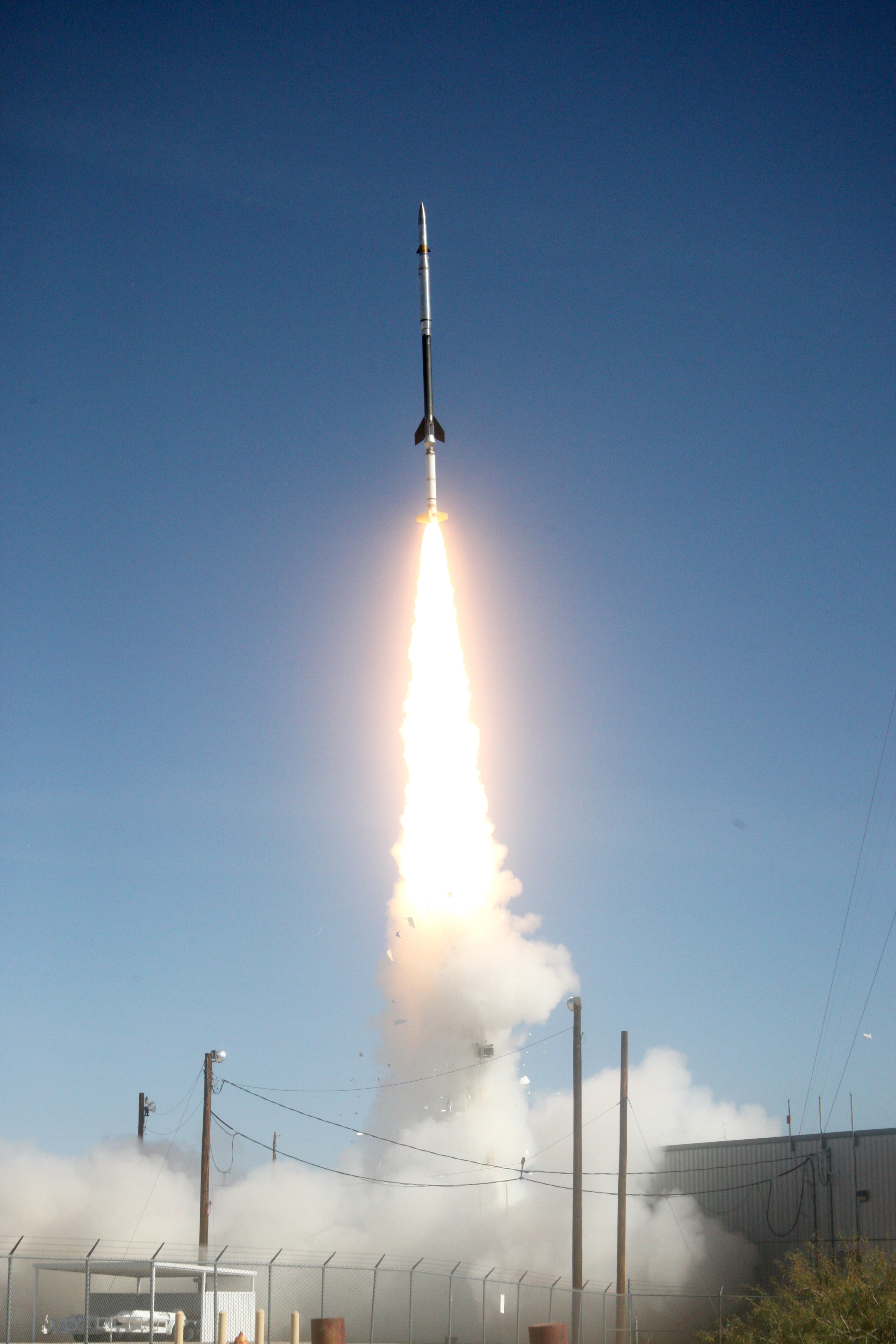}
\caption{The launch of the Focusing Optics X-ray Solar Imager (FOXSI) sounding rocket \citep{2014ApJ...793L..32K} on 2 Nov 2012 at WSMR on a Terrier-Black Brant \rom{9}. Photograph courtesy of NASA Goddard Space Flight Center and the University of California, Berkeley. }
  \label{fig:foxsi-launch-photo}
\end{minipage}
\end{figure}

Although small rockets were used by various militaries over the ensuing centuries primarily to propel artillery, their development as high altitude research tools to explore the upper atmosphere and space did not begin until the early 20th Century.  Toward this end, considerable theoretical advances were made by Russian rocketry pioneer Konstantin Tsiolkovsky (1857--1935), who advocated the exploration of space by rocket propulsion \citep{Tsiolkovsky:1903td}.  His work would influence early German and American rocket engineers, as well as help provide the foundation for subsequent Russian space exploration.

With respect to rocketry in the United States, the leading innovator of modern rocket technology and theory was the physicist Robert H. Goddard (1882--1945). A patent for his ``Rocket Apparatus'' was filed as early as 1914 \citep{goddard1914rocket}.  In his classic paper, ``A Method of Reaching Extreme Altitudes'', Goddard suggested that rockets could be used for upper atmosphere research \citep{1920Natur.105..809G}.  Pursuing that vision, Goddard developed the first liquid-fueled rocket, which he launched on 16 March 1926 \citep{marsh}. Three years later, on 17 July 1929, Goddard placed an aneroid barometer and a thermometer on one of his rockets to measure the ambient environment, unfortunately it only reached a height of 51~m.  Goddard continued his experimental launches and development of liquid fueled rockets and associated hardware throughout the 1930s.

Goddard's development and publications were of particular interest to amateur rocket groups in Germany. With the outbreak of World War II, German rocket technology took off when the Nazi government began funding the development of the `Vergeltungswaffe' (Retaliation Weapon) 1 and 2 (V-1 and V-2); the V-2 was the world's first guided ballistic missile. The V-2 stood 14 meters tall, with a diameter of 1.65 meters and a gross launch weight of about 12,700 pounds when loaded with 1,000 kg of explosives. The first successful flight of the V-2 was on 3 October 1942; the flight achieved an apogee of 84.5 km \citep{neufield}. Though their application was purely a military one, plans did exist for the scientific use of the V-2 rocket, albeit of relatively low priority and not realized during the war. The capabilities of the V-2's for high-altitude research were proven during a flight which reached an altitude of 172 km.

In parallel, the United States began its own rocket development program. In December 1944, the United States began the ``WAC Corporal'' project when the Army Signal Corps requested a high-altitude rocket ``to carry 25 pounds [11.3 kg] of meteorological instruments to 100,000 ft [30.5 km] or more.''  The WAC Corporal thus became the first sounding rocket developed in the United States designed to carry out scientific research with repeatable launches. It was first launched from White Sands, New Mexico, on September 26, 1945, well before the V-2's would be transferred to, and launched within, the United States.  The WAC Corporal would prove instrumental in the development of follow-on rockets designed and built in the United States, such as the Aerobee.

With the fall of Germany in 1945, the V-2 development team, led by Wernher von Braun, surrendered to the United States. The team, along with enough parts to build 100 V-2s, was re-located to the new White Sands Proving Ground (now the White Sands Missile Range, WSMR), where the group joined the existing WAC Corporal program to form the new Hermes Project. The Naval Research Laboratory (NRL) re-engineered the original German warhead to provide room for instrumentation, and NRL formed the Upper Atmosphere Research Panel to oversee experiments conducted with the V-2 platform. Scientific interest in the nascent sounding rocket program was already strong: a conference organized at the time included more than 50 interested scientists and engineers from over a dozen organizations. This panel would guide sounding rocket research for over a decade. 

The second US-based launch of a V-2 included instruments to measure cosmic rays as well as cameras for high altitude photography, already demonstrating the potential of sounding rockets for both astrophysics and terrestrial studies. The flight reached an altitude of 112 km. On 24 October 1946 (V-2 launch number 13), a V-2 carried a 35 mm motion picture camera with a 0.67 Hz frame rate to 100 km, where it took the first photographs of the Earth from space \citep{Holliday:1950tf}. 
The Blossom Project developed a parachute recovery system (launch 37) and also sent the first animal, a rhesus monkey named Albert I to space, opening up the field of biomedical space research \citep{Gray:nmTkcdFS}. At the same time, the Bumper Program sought to reach even higher altitudes by creating a two-stage vehicle with the WAC Corporal riding as a second stage atop a V-2 first stage.

The 50\% success rate of the V-2 launches was low compared to the standard set by modern sounding rockets over the past 20 years, with $>$90\% mission success and $>$97\% vehicle success 
\citep{1971NASSP4401.....C, NSROC:2015va}. However, 
the high altitude observing platform offered by these flights proved the importance of the sounding rocket as a tool for scientific research. Important firsts in the domain of astrophysics and solar physics from early sounding rockets include the first solar UV spectrograms \citep{1953PhRv...91..299R}, the first observations of X-rays from sources beyond the Solar System \citep{giacconi}, the first X-ray image of the Sun \citep{1963ApJ...137....3B}, and the discovery of molecular hydrogen in the interstellar medium \citep{carruthers}.
In addition, these flights provided an important experience for the burgeoning fields of sounding rocket science which enabled the development of new technologies such as telemetering techniques, recovery, and instrument pointing.  Scientists also learned how to build instruments which could survive the intense vibration of a rocket launch and fit inside the small available volume. 

Following on this experience, the US embarked on the development of numerous liquid- and solid-fueled sounding rockets including the Viking, Aerobee, and Nike-Deacon, among others.  These rockets would become important elements of various research programs across the country, eventually providing the foundation for the NASA Sounding Rocket Program Office (SRPO), which has been a key element of space research and exploration at NASA since the agency's inception in 1958. The importance of sounding rockets only grew with the development of satellite observatories, adding the role of low-cost proving ground for these more expensive space missions.




%% file: sec-capabilities-rockets.tex

The number and types of sounding rocket motors available to the science community since the earliest days of NASA have been numerous.  The choice of vehicle for a given science experiment depends critically on the desired apogee (peak altitude) and range requirements as well as the payload mass, diameter, and length. 

The primary vehicle used by the solar and astrophysics communities is the solid-fuel Terrier-Black Brant \rom{9} (see \S~\ref{sec:science-platform}). This two-stage vehicle is a member of a family of vehicles which use the 0.44 m diameter, 5.6 m long Black Brant \rom{5} motor. The Terrier-Black Brant \rom{9} adds to that a Terrier MK first stage (0.46 m diameter and 4.3 m long), providing experimenters with excellent performance in apogee versus payload mass and typically achieving 400 km apogees for a payload mass of 360 kg. Prior to adding the payload, this two-stage rocket weighs 1,290 kg which includes 1,000 kg of solid propellant. The Black-Brant motor provides a thrust of 81,000 Newtons over 34 s while the Terrier first stage provides 89,000 Newtons of thrust over 6 s. This configuration permits payload sections of 18.5 inches or 22 inch diameter and lengths of 2.54 m to 8.9 m. A photograph of a day-time launch of the FOXSI \citep{2014ApJ...793L..32K} payload by a Terrier-Black Brant \rom{9} is shown in Figure~\ref{fig:foxsi-launch-photo}.

\begin{figure}
\begin{minipage}{0.45\linewidth}
\includegraphics[width=2in]{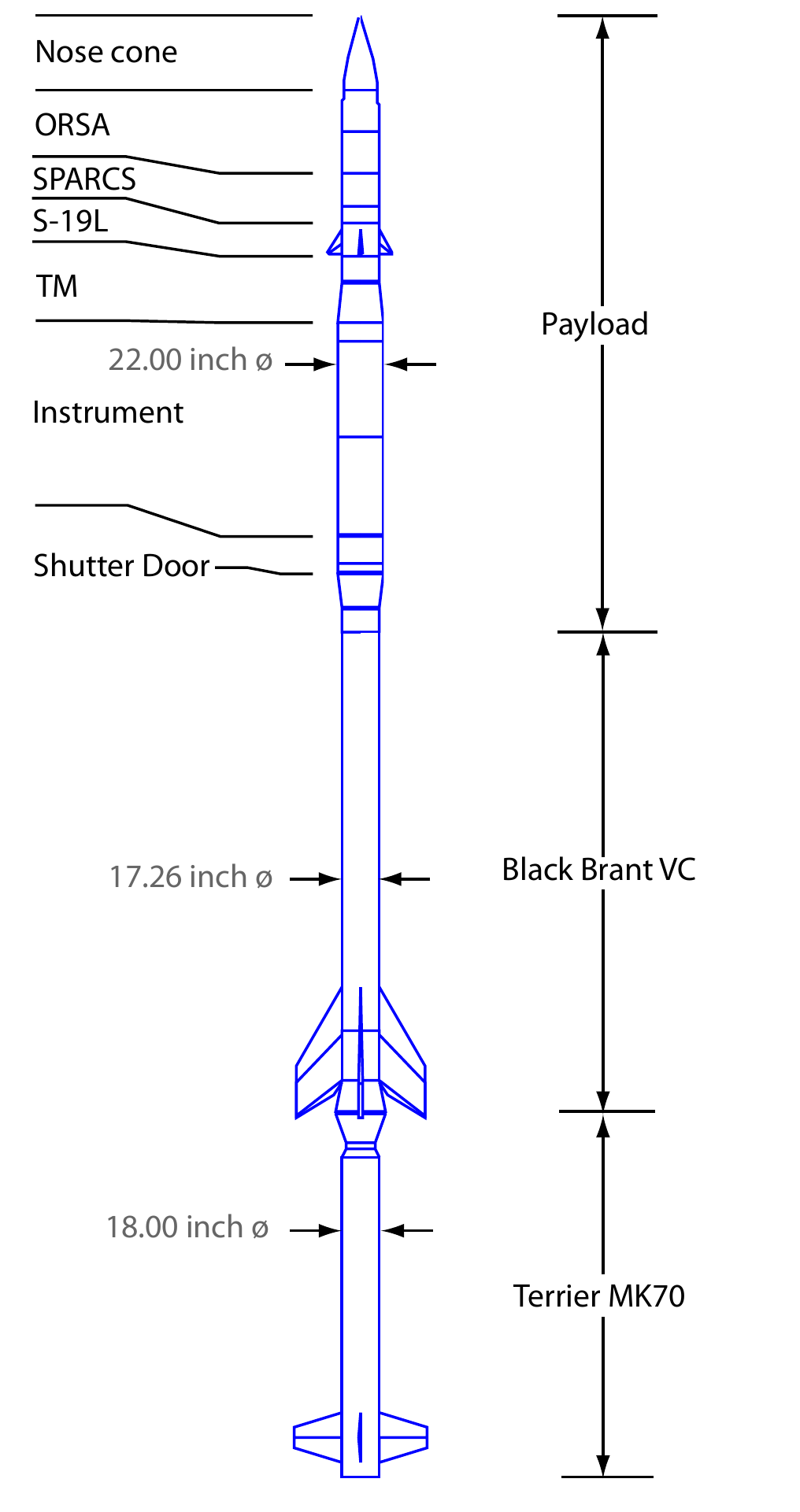}
\caption{A diagram of a typical Terrier-Black Brant \rom{9} sounding rocket carrying a solar payload. In an astrophysical one, the SPARCS section is replaced by components of the CACS. The sensors for the SPARCS or CACS systems are located within the experiment section itself to assure alignment with the telescope.}
\label{fig:rocket-diagram}
\end{minipage}%
%
\begin{minipage}{0.45\linewidth}
\vspace{-4cm}
\caption{Typical Black Brant \rom{9} Launch Timeline for a Solar Payload}
\begin{tabular}{@{}lrr@{}} \toprule
Event & Time & Altitude \\
 & (s) & (km) \\
\colrule
Motor Ignition & 0 & 0 \\
Terrier Burnout & 6 & 3 \\
Black Brant Ignition & 12 & 6 \\
Black Brant Burnout & 46 & 50 \\
Despin & 60 & 80 \\
Payload Separation & 64 & 88 \\
SPARCS/CACS Enable & 66 & 93 \\
Shutter Door Open & 67 & 95 \\
Observations Begin & 110 & 180 \\
Apogee & 300 & 330 \\
Shutter Door Closes & & \\
Observations End & 500 & 140 \\
Parachute Deploy & 690 & 2 \\
Payload Impact & 1000 & 0 \\
\botrule
\end{tabular}
\label{tab:launch-timeline}
\end{minipage}%
\end{figure}

A typical payload consists of the experiment as well as a number of sub-systems provided by the
engineering team at WFF. A typical solar configuration (in this case the FOXSI payload) is shown in Figure~\ref{fig:rocket-diagram}. The payload consists of the following sections, (1) the nose cone, (2) the Ogive Recovery System Assembly (ORSA) which includes the parachute, (3) the Solar Pointing Attitude Rocket Control System (SPARCS) for solar instruments or Celestial Attitude Control System (CACS) for astrophysical experiments, (4) the S-19L Boost Guidance system, (5) the Telemetry (TM) section, (6) the experiment, and (7) the shutter door.

A typical flight timeline for a Black Brant \rom{9} rocket carrying an instrument is given in Table~\ref{tab:launch-timeline}. Rocket burnout occurs 46 s after launch. The rocket is spun up for stability during ascent. Despin occurs at around 60 s and payload separation seconds later. Soon thereafter, the fine pointing system (either SPARCS or CACS) is enabled and begins target acquisition, and the shutter door opens to expose the experiment (for an astrophysical payload, with the ST-5000 star camera inside the experiment section, the door/pointing order is reversed). Observations can begin once the pointing system provides fine pointing control (110 s). Observations end once the shutter door closes at 400--500 s (depending on apogee). Power to the experiment is cut one minute later. The parachute is deployed at 12 minutes after launch and the payload lands approximately 5 minutes later for a total time in the air of $\sim$17 minutes. Recovery, if available, takes place within a couple of hours after landing.

%% file: sec-capabilities-launch-sites.tex

\begin{figure}
\begin{center}\includegraphics[width=6in]{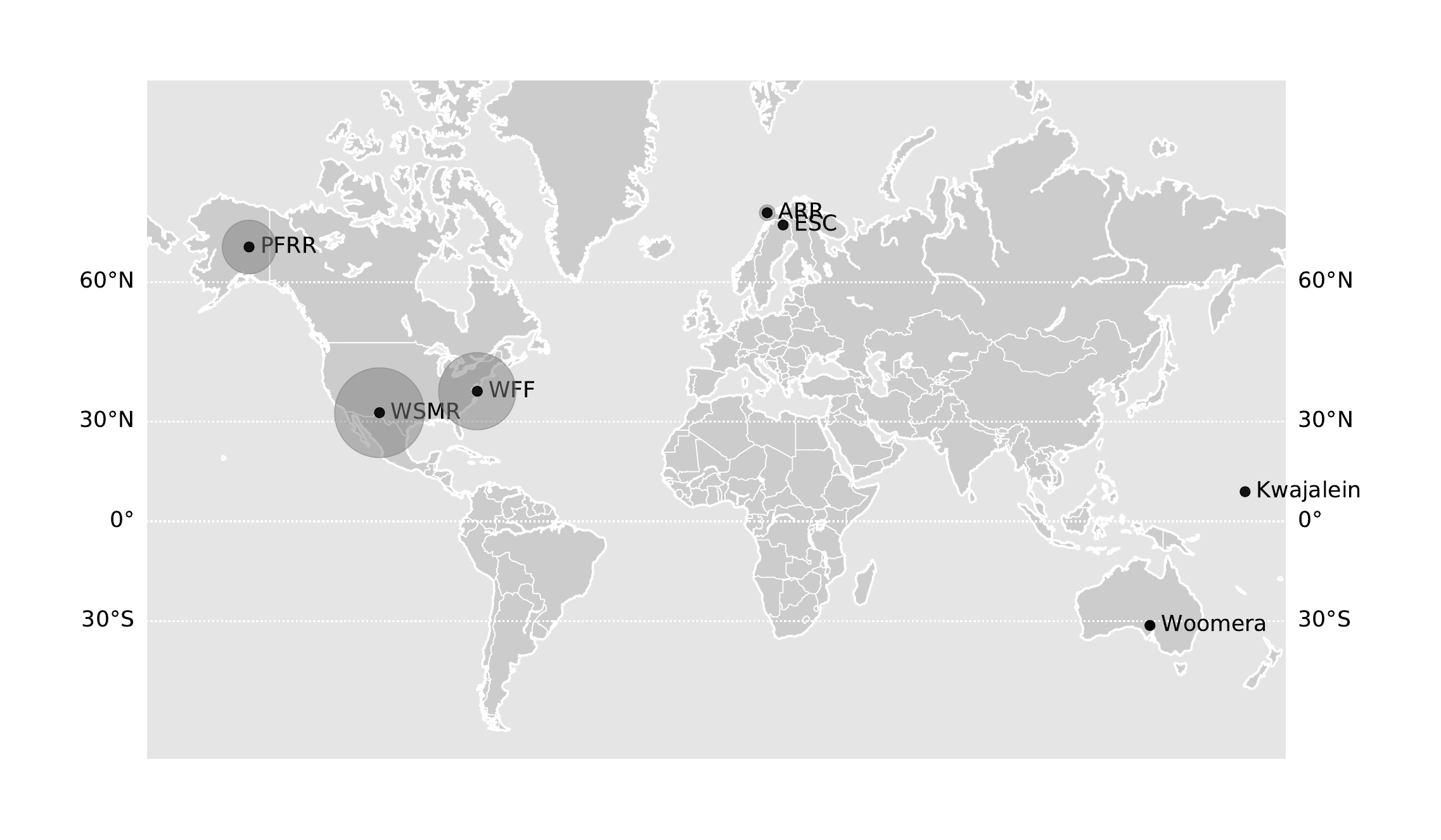}\end{center}
\caption{
A map showing the locations of the established NASA launch sites.  The size of the circle corresponds to the relative number of NASA launches in the last $\sim$11 years. A few sites have not hosted any NASA launches in the period considered here though they are available launch sites including Esrange Space Center and Woomera.  Kwajalein hosted 2 launches.}
\label{fig:launch-site-map}
\end{figure}

NASA sounding rockets are launched from a number of different locations, including established ranges and temporary mobile ranges.  The established ranges in use by the NASA program are shown in Figure~\ref{fig:launch-site-map} and listed in Table~\ref{tab:launch-sites}. 

\begin{wstable}[h]
\caption{Established Launch Sites Currently Used by NASA}
\begin{tabular}{@{}llrrlll@{}} \toprule
Name & Nickname & Latitude & Longitude & Location & Payload Recovery & \# of launches\\
\colrule
Wallops Island, VA & WFF & 37.94028 & -75.46639 & VA, U.S. & No, in dev.(water) & 51 \\
Poker Flat Research Range & PFRR & 65.12525 & -147.48802 & AK, U.S. & No & 34 \\
White Sands Missile Range & WSMR & 32.41780 & -106.32087 & NM, U.S. & Yes & 60\\
And\o ya Space Center & ARR & 69.29508 & 16.02999 & Norway & Yes (water) & 12\\
Esrange Space Center & ESC& 67.89304 & 21.06489 & Sweden & Yes & 0 \\
Kwajalein Atoll & Kwajalein & 9.31227 & 166.87474 & Marshall Islands  & No & 2\\
Woomera Test Range & Woomera & -31.20145 & 136.83385 & Australia & Yes & 0 \\
\botrule
\end{tabular}
\label{tab:launch-sites}
\end{wstable}

All sites excepting Woomera (Australia) are located in the Northern Hemisphere. Accordingly, Woomera is often used to facilitate important astrophysical observations of the Southern sky. Additional launch sites are available with sites in Norway (And\o ya Space Center) and Sweden (Esrange Space Center), providing access to higher latitudes than the Poker Flat Research Range (PFRR) as well as recovery services. Kwajalein Atoll, in the Marshall Islands, provides access to near equatorial latitudes.


%% file: sec-capabilities-pointing-sparcs.tex

NASA's time-tested SPARCS pointing systems is the primary means to point a payload at the Sun. It consists of three optical sensors (Coarse Sun Sensor - CSS, Miniature Acquisition Sun Sensor - MASS, Lockheed Intermediate Sun Sensor - LISS), two gyroscopic systems, and one magnetometer.  Control is performed through a reaction control system using cold-gas pneumatics. The CSS can point the payload at the Sun from any orientation before the finer sensors take over. The pointing accuracy of the system in pitch and yaw is better than 40 arcsec with a stability of $\sim$0.1 arcsec. The roll orientation accuracy is around 2$^\circ$ with a high stability ($<$ 0.04$^\circ$/hr) and low jitter ($<$0.01$^\circ$). Pointing knowledge in pitch and yaw is provided by the MASS \& LISS sensors that both consist of quad photocells. The difference in output across the horizontal and vertical sensors provides a signal with X and Y amplitudes proportional to the offset from the target. The LISS provides the highest accuracy, with a typical stability limited by the digital-to-analog converter of 0.05 arcsec\footnote{Jesus Martinez, private communication}.

A significant advantage of this system is that it provides the capability to repoint the payload during the flight based on real-time observations. Slewing to a new target takes about 2 to 4 s. The overall pointing accuracy of the payload is generally limited by the alignment between the LISS and the telescope.

%% file: sec-capabilities-pointing-star.tex

Many, but not all, astrophysical payloads involve telescopes and most make use of the CACS for targeting and control. The CACS is distributed across multiple sections of the payload, with the ST-5000 star tracker, developed at the University of Wisconsin \citep{st5000}, located behind the shutter door and within the payload to allow precise alignment, the GLN-MAC fiber optic gyro (for coarse guidance when pointing near the horizon and orientation at beginning of flight) and electronics deck are located forward of the center of gravity, and the pitch and yaw gas tank and nozzles are located near the ORSA section at the forward end for maximum torque.

The Celestial ACS provides absolute pointing accuracy to 2--5 arcmin. Some payloads have achieved sub-arcsecond pointing with additional guidance either on-board or using the video downlink and the command uplink system (\S\ref{sec-capabilities-telemetry}). Stability is around 10 arcsec s$^{-1}$. Additional control modules (each of which adds mass to the payload) can reduce the jitter to 2--3 arcsec s$^{-1}$ or down to $<$1 arcsec s$^{-1}$, depending on the needs of the payload. The PICTURE payload, with additional internal control of the optics, demonstrated pointing stability at the marcsec \citep{pictureb}. Heavier payloads can also necessitate the addition of up to two additional gas pressure vessels for gas for inertial control, and the penalty for higher mass and lower apogee and flight time.

Slew times between targets are typically $\sim$20 s, with only minor increases for long slews since the majority of time is spent settling on a target. The ST-5000 is capable of orienting itself from anywhere in the sky within 7 s, but beginning of life initialization typically takes 40 s. 

%% file: sec-capabilities-telemetry.tex

Transmitting data during flight is the primary means of accessing data from the payload, although many payloads employ on-board storage as a backup.  SRPO has a telemetry system with both down-link and up-link capability.  Down-link operates principally through S-Band (2200--2395 MHz) and routinely provides data rates of 10 Mbit/s.  Over a 5-6 min flight this corresponds to a total data volume of 3--3.6 Gbit. Data transmission systems which include digital telemetry techniques such as Bi-phase Pulse Code Modulation can provide data rates up to 20 Mbit/s.  Shaped-offset quadrature phase-shift keying (SOQPSK) options are expected to increase these rates in the future.  Real-time analog TV camera NTSC video signals can easily be provided while high-definition HDMI will soon be available.

Solar and astrophysical payloads typically include uplink command capability to enable the scientist to actively point the telescope to different targets during the flight or command the experiment.  The command uplink system operates at 437.5 MHz and uses FSK modulation.  Uplink command rates are considerably slower than downlink data transfer, but since these are manual commands, ``several commands per second'' \citep{NSROC:2015va} are usually sufficient.

%% file: sec-capabilities-recovery.tex

Solar and astrophysics payloads are usually expensive and complex, and also require multiple years to build. Accordingly, payload recovery is a typical requirement, which is why WSMR is a popular launch location. Recovery enables reuse of the payload, including the instrument, the fine-pointing system (CACS/SPARCS), and other sub-systems.  Another advantage is that, in some cases, recovery enables scientific data to be retrieved, particularly in cases where experiments generate large data volumes that cannot be transmitted in real-time and are stored in on-board memory.  The combination of a parachute and a crush bumper provide the payload with a soft landing. Land recovery is usually performed by helicopter, in which the payload is lifted via a hoist cable and returned as a unit. 

NASA has recovered several payloads launched from Poker Flat, but most payloads launched from there are not recovered. Both WFF and ARR provide water recovery for non-telescope payloads with relatively low apogees (generally less than 250 km).  This is provided via boats, although ``in air'' snatches of light payloads have been demonstrated at WFF.  WFF is currently exploring means to recover telescope payloads which will enable higher altitude rockets and more observing time than is currently possible from WSMR (see \S\ref{sec:future})

%% file: sec-science-platform.tex

In order to illustrate how sounding rockets are used within NASA's program, we examined information for launches from the period of 2005 to the end of 2015 provided by SRPO, a total period of approximately 11 years. This data set includes a total of 159 launches in four science disciplines:  solar, astrophysics, geospace, and ``other'' including test launches, student outreach, and special projects. Reimbursable launches (launches conducted by SRPO but whose payloads are not funded by NASA) are not included in this data set.

Every launch includes information such as the launch category, the principal investigator (PI) and PI institution, and launch site. Many launches
have further information such as the mission name (135 launches), apogee (150 launches), impact range (142 launches), 
and launch vehicle (153 launches).

\begin{figure}
\begin{center}\includegraphics[width=4in]{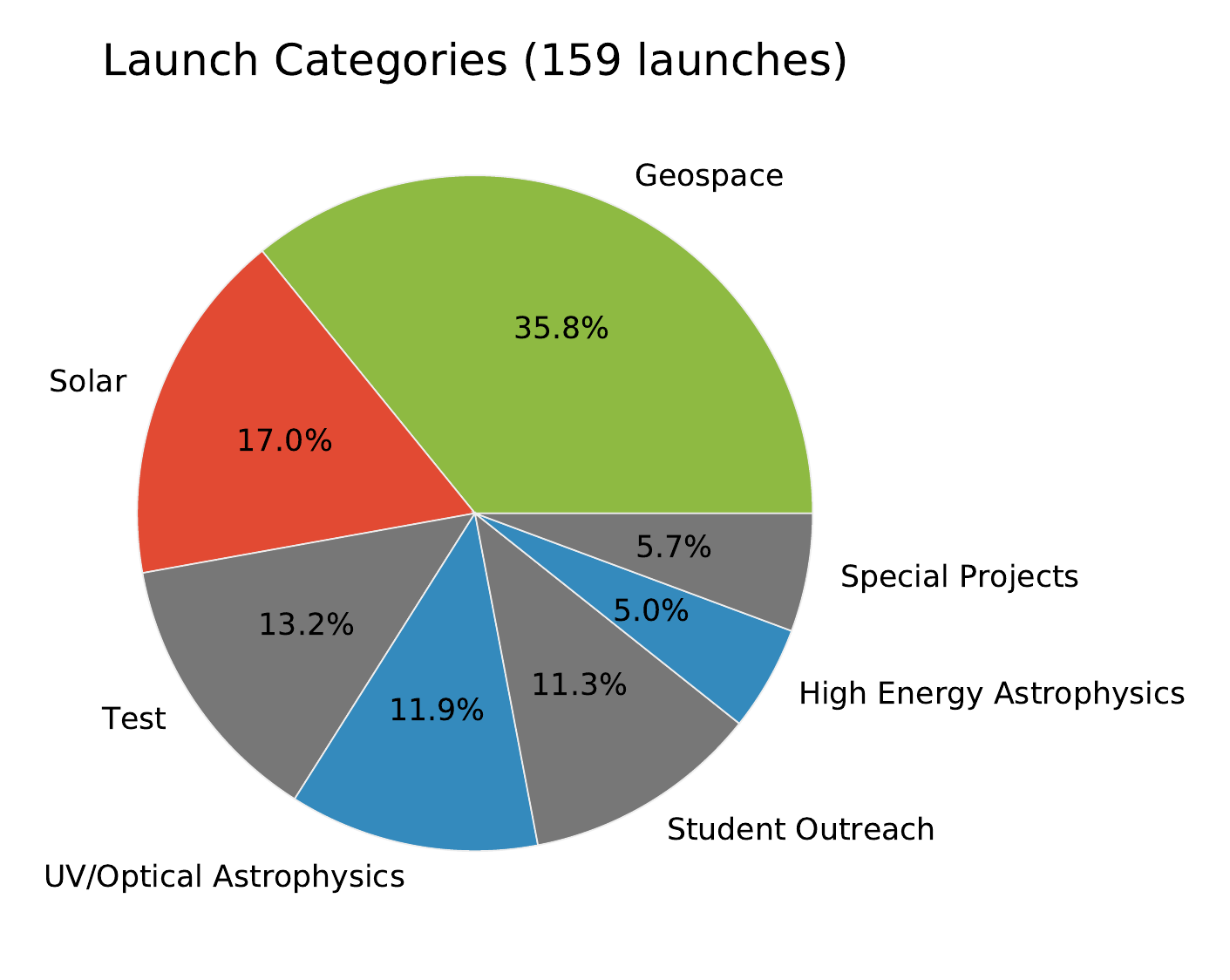}\end{center}
\caption{A pie chart showing the breakdown of 159 sounding rocket launches by category from 2005 and 2016.}
\label{fig:launch-categories}
\end{figure}

\begin{figure}
\begin{minipage}{.45\textwidth}
\centering
\includegraphics[width=1.0\linewidth]{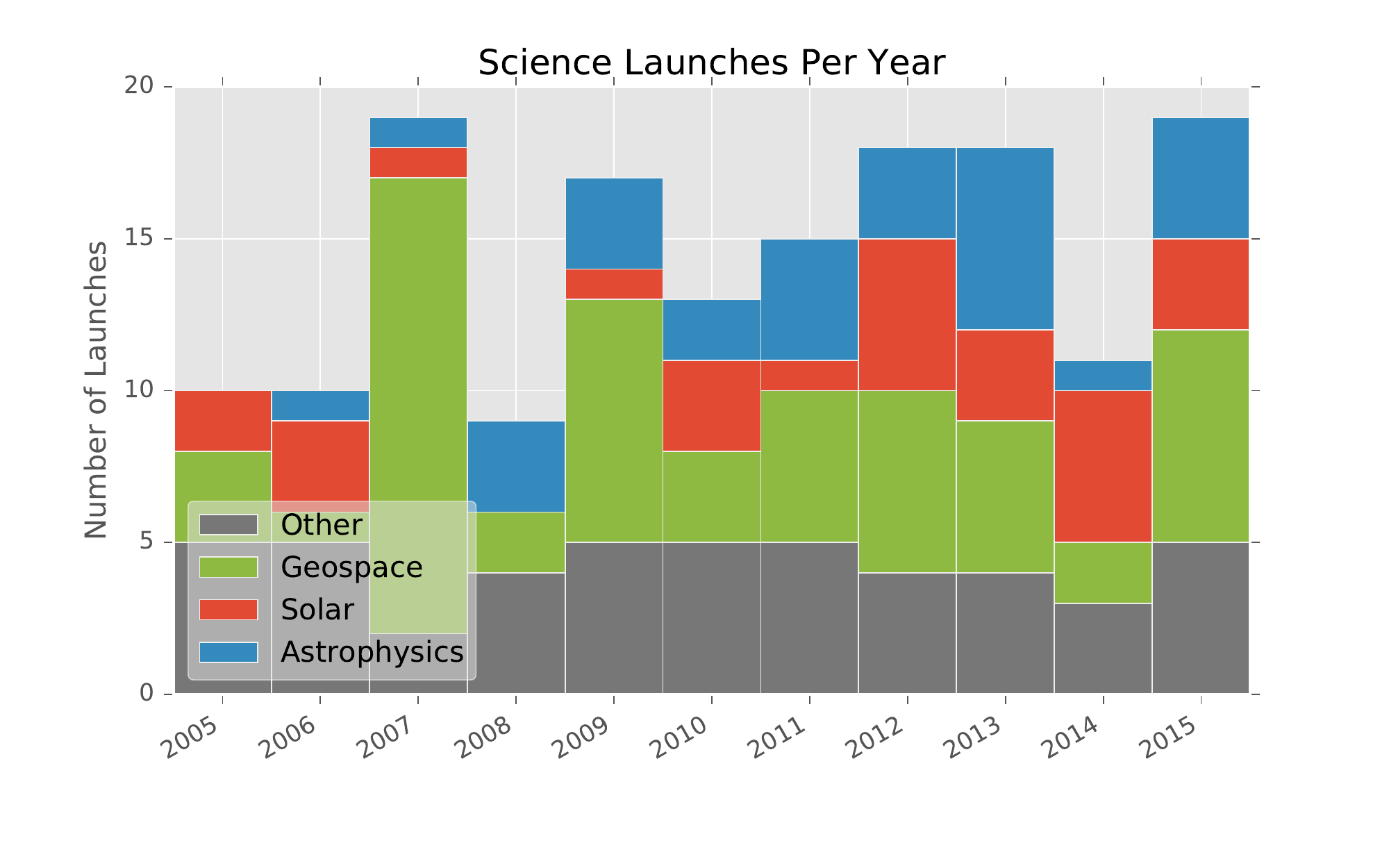}
\caption{The distribution of NASA sounding rocket launches as a function of time.}
\label{fig:launch-peryear}
\end{minipage}\hspace{0.01\textwidth}
\begin{minipage}{.45\textwidth}
\centering
\includegraphics[width=1.0\linewidth]{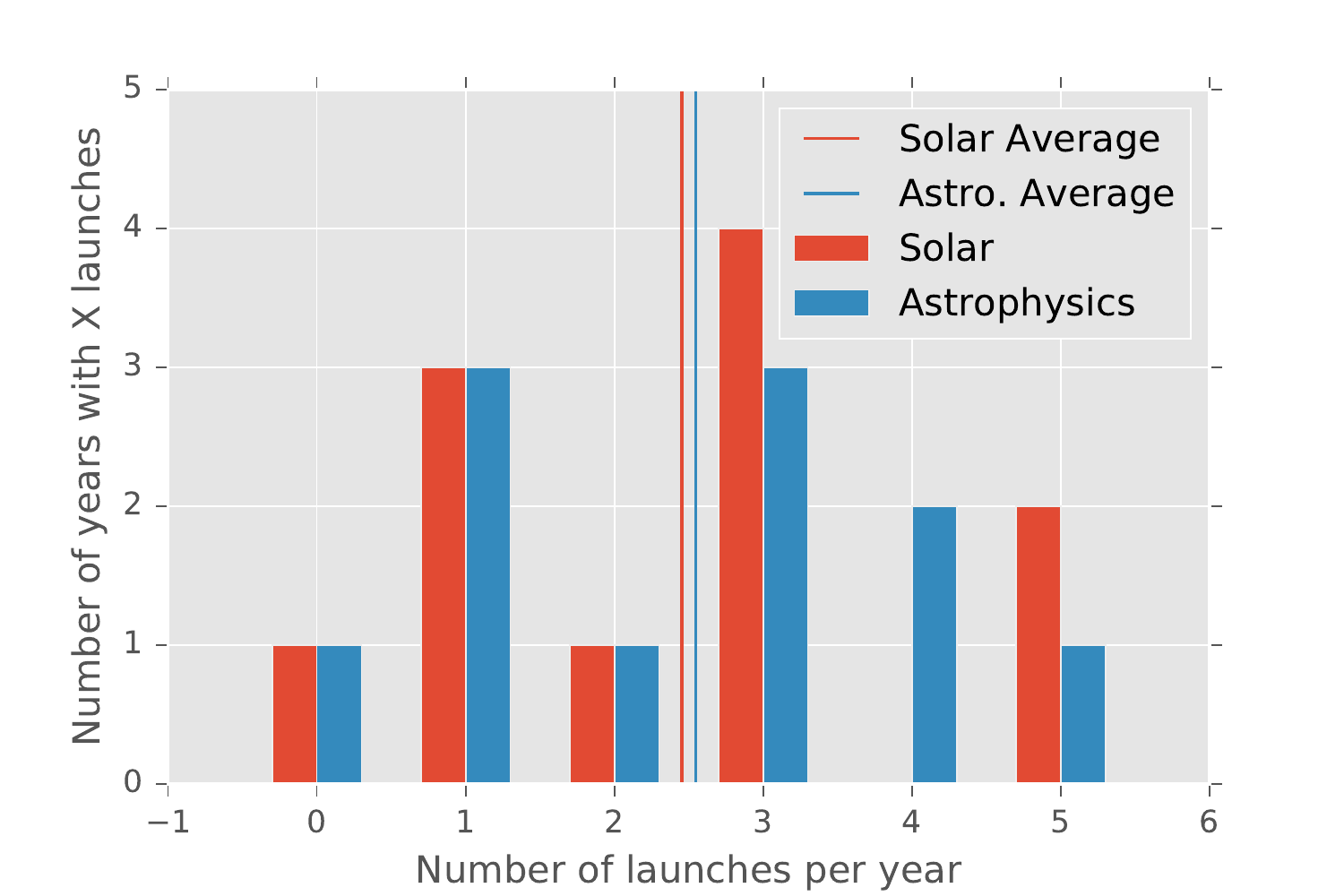}
\caption{The distribution of launch frequency per year. The vertical lines show the average number of launches per year ($\sim$2.5 launches per year for solar and astrophysics).}
\end{minipage}
\label{fig:launch-peryear-hist}
\end{figure}

Figure~\ref{fig:launch-categories} provides a breakdown of all of the launches by category.  The majority of launches are devoted to geospace (57 launches) while 55 of the launches were in the solar or astrophysics categories, for a total of 112 science launches. A total of 27 launches  had the Sun as their target while 28 launches pointed to astrophysical targets. The remainder of the launches (48 launches) were devoted to other projects such as testing and development or student outreach. Averaged over the total time period this implies an average of $\sim$5 solar and astro launches per year or 2.5 launches each. Figure~\ref{fig:launch-peryear} shows the distribution of launches as a function of time and Figure~\ref{fig:launch-peryear-hist} show a histogram of the number of launches per year. Only 1 year had no launches for either solar (2008) or astrophysics flights (2005).

\subsection{Users}

\begin{figure}
\begin{center}\includegraphics[width=5in]{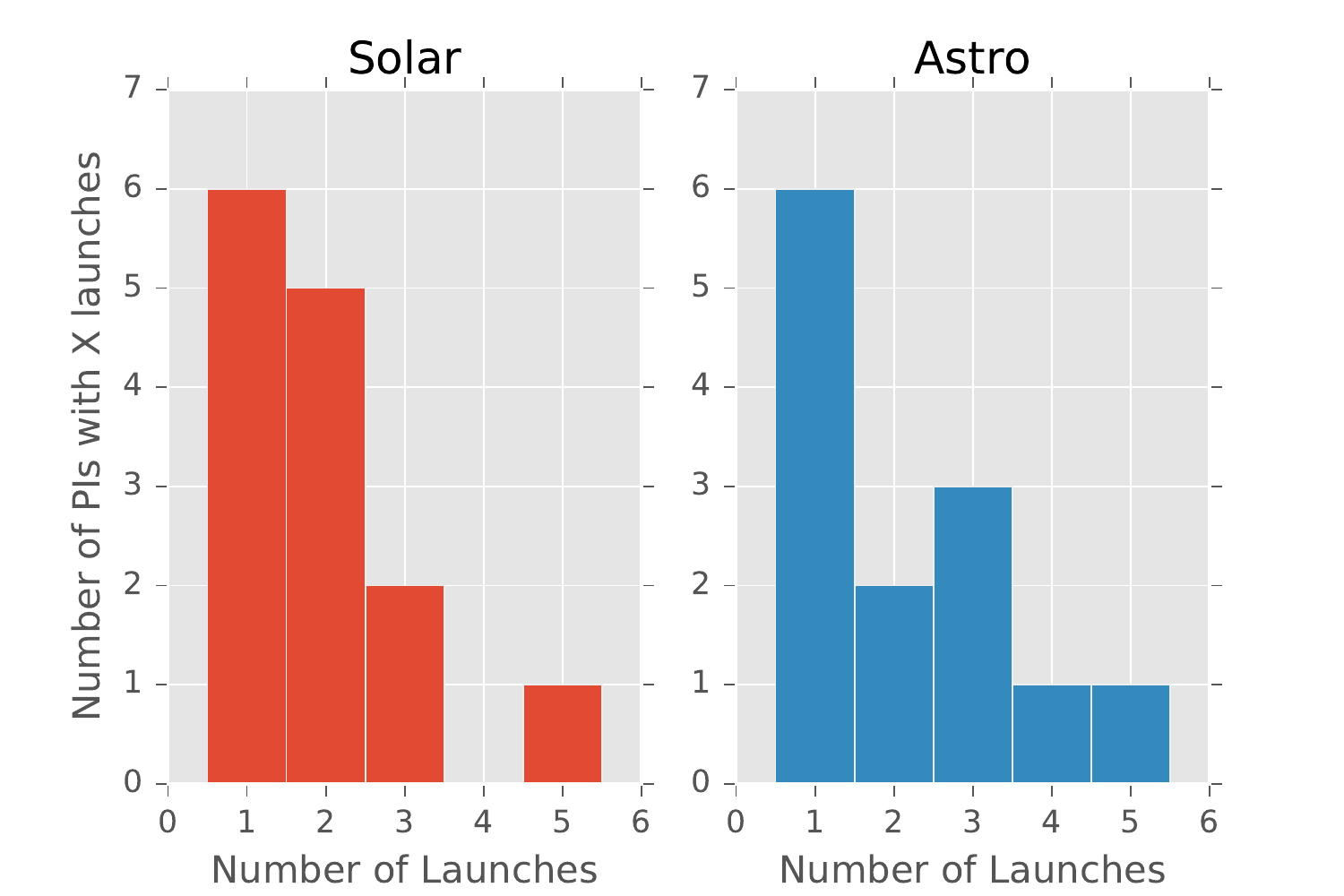}\end{center}
\caption{The distribution of launches per user.}
\label{fig:launch_perpi}
\end{figure}

Considering only solar launches, there are 14 PIs and 9 institutions associated with 27 launches. The average is $\sim$2 launches per PI and $\sim$3 launches per institution. Figure~\ref{fig:launch_perpi} shows the distribution of number of launches per PI. The maximum number of launches for a single PI was 5 though this represents a series of calibration flights (see \S\ref{sec:sdo-eve}). Two had 3 launches each while 22\% of launches are from PIs with only one launch. The most prolific institutions in terms of solar launches had 5 launches from a single PI followed closely by two institutions each with 4 launches. These three institutions represent 52\% of all solar launches. NASA centers had about a quarter of solar launches, with a total of 7 launches.

For astrophysics, the statistics are similar. There are 13 individual PIs and 9 institutions associated with 28 launches. The average is then 2.1 launches per PI and 3 launches per institution. The maximum number of launches for a single PI is 5. Prolific PIs (those with three or more launches) represent a majority (64\%) of all launches. For comparison, 21\% of PIs had only had one launch. The institution with the most launches has 8 launches (28\% of astrophysical launches) while the next-most prolific institution has only one more than half that number. Combined, these two institutions represent almost half of all astrophysical launches.

The average time required to develop a new sounding rocket payload is 3-5 years (sometimes longer). To upgrade and prepare a re-flight is approximately 2 years. The maximum number of flights for a single payload in the $\sim$11~yr studied is 4 (assuming a new payload) to 5 (assuming only reflights during the time period), although most payloads will not fly so frequently. Though the sample size is small the distributions for solar and astro launches suggests a mix of mature and new payloads with some reflights. 



\subsection{Scientific Impact}
\label{science-impact}
The scientific impact of any field is a difficult metric to define. We make a coarse estimate by searching NASA's Astrophysics Data System for refereed articles with abstracts or titles that include payload names or mission numbers as reported in the SRPO launch database. This catches articles written by collaborators but misses articles that only use phrases such as ``data from a sounding rocket flight’’ or reference the flight in the article but not in the abstract. It also excludes sounding rocket publications regarding pre-2005 payloads even if published during this window and as-yet-unflown payloads that exist and might have related instrumentation papers, and it misses unrefereed conference proceedings such as {\it Proceedings of SPIE}, where many optical and instrument results are reported. Additionally, this method misses papers where rocket names differ from names used in the publication (including the PhD thesis of one of the authors of this paper).

From 01 Jan 2005 to 20 Oct 2015, 55 refereed articles have been published about astrophysics and solar (not geophysics) payloads meeting these criteria, with 37 solar and 18 astrophysical. Collectively, they have been cited 568 times, for an average of 10.3 citations per paper. The citation statistics are skewed heavily toward solar payloads, with 530 to the 38 astrophysics citations. The EVE and Hi-C solar payloads (\S \ref{sec:solar-science}) dominate these statistics, with 21 publications and 359 citations between them; both of these payloads are discussed further in \S \ref{sec:solar-science}. Both publication and citation data are conservative estimates of the true publication impact of the payloads, since the sample contains only recent flights where data might still be under analysis or the resulting publications are too recent for their impact to be fully realized by a metric based on the number of citations.

On the educational side, 10 PhD theses were published from these payloads (by the same search criteria), with 3 solar and 7 astrophysics. As with citation and publication data, the search criteria make this a highly conservative estimate --- training of graduate students is an explicit goal of the suborbital program, and the actual ratio of rocket flights to related PhD theses is, anecdotally, probably closer to unity.



\subsection{Launch Site Usage}
To facilitate the wide range of scientific targets pursued by sounding rocket missions, NASA supports a number of different launch sites across the world (see \S\ref{sec-capabilities-launch-sites}). These include a number of successful geospace campaigns at `mobile' sites, for example in Greenland, Brazil, Peru, and Puerto Rico, as well as astrophysical campaigns in Australia that permit viewing of the southern hemisphere sky. There were no such mobile launches during the period considered here, although at least one campaign is currently being planned. The most-used established sites for science payloads are located in the continental United States (WSMR: 60 launches; WFF: 51 launches) and Alaska (PFFR - 34 launches).  Foreign sites show much less usage with ARR 
at 12 launches and no NASA launches from the Swedish Esrange Space Center or Woomera. Kwajalein had 2 launches. Solar and astrophysics launches occur almost exclusively at WSMR. Only two solar/astrophysics launches did not take place at WSMR (PFRR, when extenuating circumstances prevented launches from WSMR, and WFF for very high apogee for long observing time). This is not surprising, as WSMR permits land recovery and relatively high apogees (400~km for some payloads, affording about 6-7 minutes of observations).


\subsection{Launch Vehicle and Performance}
\begin{figure}
\begin{center}\includegraphics[width=5in]{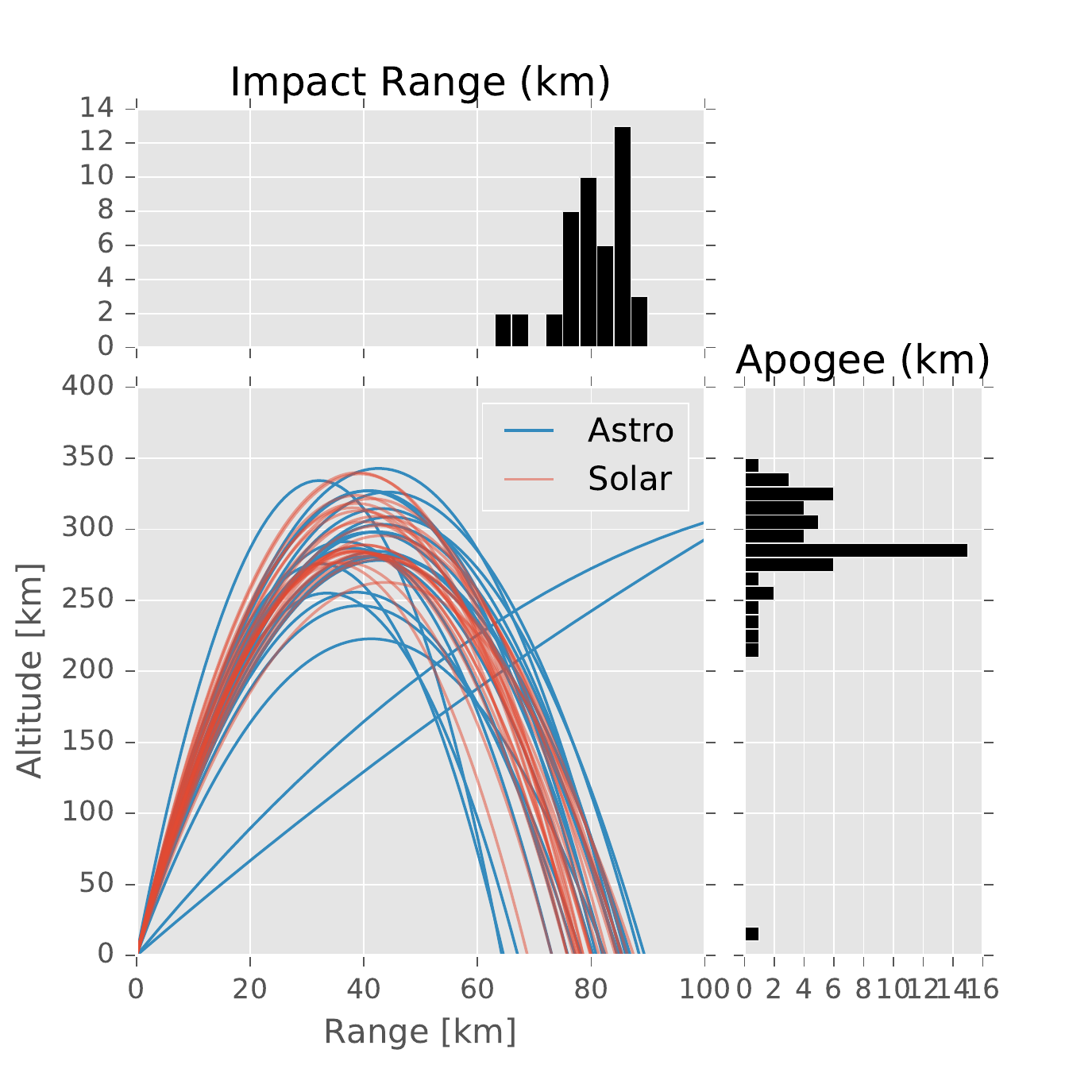}\end{center}
\caption{Launch vehicle performance for solar and astrophysics launches. Almost all launches took place at the WSMR and used a common launch vehicle (the Black Brant \rom{9}). Profiles have been idealized for a parabolic trajectory with the correct apogee and down-range distance, while a real profile would be a narrower profile with a tail drifting down-range after the parachute opens on descent. The two outlier profiles show astrophysics launches, one of which took place from PFRR with an apogee similar to the flights from WSMR.  Another rocket flew from WFF with an apogee of 577 km that did not include recovery. This rocket flight included over 10 minutes of observing time. One solar launch with a small apogee (10 km) is represented which was due to an early flight termination.
}
\label{fig:launch-performance}
\end{figure}

The choice of launch vehicle for remote sensing applications such as solar and astrophysics is mostly driven by observing time and altitude. Astrophysicists and solar physicists are therefore heavy users of the Black Brant \rom{9} rocket engine (\S\ref{sec-capabilities-rockets}). Because of the common choice of launch vehicle, the flight performance is very similar across payloads. Figure~\ref{fig:launch-performance} shows simulated launch trajectories for the combined astrophysics and solar rockets launched from WSMR during the 2005 to 2015 period. 
The mean apogee for solar launches is 286~km and for astro launches it is 297~km.
The impact range is constrained by the geography of the White Sands Missile Range with a mean of approximately 80~km. In contrast, one astrophysics rocket was launched from Wallops Flight Facility that did not include recovery.  This rocket was launched on a four-stage Black Brant \rom{12} vehicle which achieved an apogee of 557 km and provided over 10 minutes of observing time, enabling the PI to observe 11 different targets.

%% file: sec-solar-payloads.tex

\subsubsection{The High-resolution Coronal Imager (Hi-C)}
Hi-C was launched on 11 July 2012 and took images of the 1.5 MK solar corona with a resolution of 0.2 arcsec ($\sim$150 km), 6 times better than that available from the Atmospheric Imaging Assembly (AIA) instrument aboard the Solar Dynamics Observatory (SDO) satellite \citep{Lemen:2011jg}. The Hi-C instrument imaged the Sun in a single narrow (5 \AA) passband in the extreme ultraviolet (193 \AA) which is dominated by emission from Fe XII. With a 5.5$\times$5.5 arcmin field of view, the Hi-C instrument collects images with a 5.4 s cadence using a Ritchey-Chr\'etien telescope and a liquid nitrogen cooled CCD camera. Hi-C was flown from WSMR, reached an apogee of 267 km, and was recovered. 

The high resolution observations by Hi-C showed evidence of braided magnetic fields at several sites in the field of view \citep{2013Natur.493..501C}. Simultaneous observations by AIA in the same passband at a resolution of 1.2 arcsec could not resolve this. These features are likely the sites of the build-up of magnetic energy by convection-driven motion in the photosphere. Their subsequent relaxing to a lower energy state through magnetic reconnection may be the primary energy source for the heating of the corona.

\subsubsection{Extreme Ultraviolet Normal Incidence Spectrograph (EUNIS)}
\label{sec-eunis}
\begin{figure}
\begin{center}\includegraphics[width=6in]{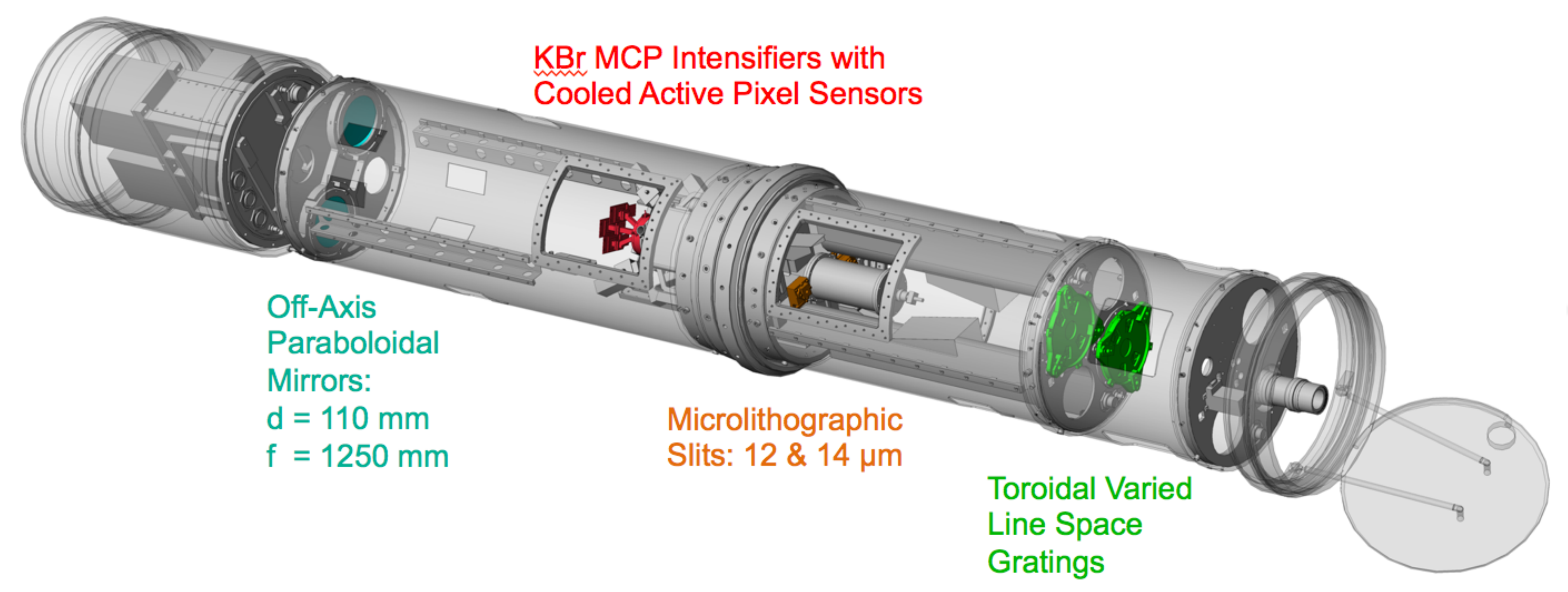}\end{center}
\caption{EUNIS comprises two co-aligned EUV imaging spectrographs.  For each
channel, light enters from the left, is imaged by an off-axis
paraboloidal primary mirror (teal) on to a slit (orange-brown) which
passes light to a toroidal grating (green) that focuses a spectrally
dispersed image on to a KBr photocathode deposited on to a Microchannel
Plate (MCP) intensifier tube that is fiber-optically coupled to three
cooled active pixel sensors (red).  The slits are 660 arcsec long
with side lobes to assist with co-alignment to data from other
instruments.  For the 2013 flight, the optics for the 300-370 A channel
were coated with an aperiodic Ir/Si multilayer, and those for the
525-635 A channel were coated with B$_{4}$C/Ir.}
\label{fig:eunis}
\end{figure}

The Extreme Ultraviolet Normal Incidence Spectrograph (EUNIS) was launched on 23 April 2013 from WSMR. This payload had flown twice before but was significantly upgraded for this latest flight \citep{2014ApJ...790..112B}. It made new spatially resolved EUV spectroscopic measurements of the Sun using a two-channel imaging spectrograph to observe the 300--370 \AA\ and 525--635\AA\ wavelength bands with spatial resolutions of $\sim$4 arcsec and 3 arcsec, respectively, and the best dynamic range ever achieved in this range. The 2013 flight incorporated a Toroidal Varied Line Space grating for the new 525--635 \AA\ waveband, and provided the first flight demonstration of cooled ($-20^\circ$C) Active Pixel Sensor (APS) detectors (see Figure~\ref{fig:eunis}).  During it's flight, it observed emission from a faint line from Fe \rom{19} at 592.236 \AA. This emission line has a formation temperature of 8.9 MK. This temperature is frequently associated with flare-heated plasma but no macroscopic flaring was seen at the time of this observation. It is therefore interpreted as a likely signature of small and pervasive impulsive heating events (so-called nanoflares) which have been theorized \citep{1975ApJ...199L..53G, 1983ApJ...264..642P, 1988ApJ...330..474P, Lin:1984cf, Cargill:1994jt, Klimchuk:2008cn} as the solution to the solar coronal heating problem \citep{1939NW.....27..214G, Edlen:1943uj}. The line emission was observed over a large portion of active region AR 11726 suggesting that it is not related to macroscopic flaring. Additionally, no significant doppler shifts were observed in this line indicating that the Fe \rom{19} was not from plasma propelled upward but is most likely produced by many localized impulsive heating events.

\subsubsection{SDO EVE Calibration Rocket}
\label{sec:sdo-eve}
The primary purpose of this mission is to provide underflight calibration of the Extreme UV Variability Experiment (EVE, \citet{2012SoPh..275..115W}) aboard the SDO satellite \citep{Pesnell:2011ik} and to several other EUV instruments such as SWAP and LYRA on Proba2\citep{Didkovsky:2012us}. 
Space-based EUV instruments frequently suffer degradation over their mission lifetime from affects as exposure to charged particles. Their sensitivity must therefore be checked often. Sounding rockets payloads provide a unique capability for doing this because don't degrade during to their short exposure to the space environment, and can be accurately calibrated on the ground both before and after flights. Calibrations flights have so far taken place roughly every year and a half starting on May 3, 2010. The latest flight was planned for May 2015. During the last two flights (23 Jun 2012, 21 October 2013), a new soft x-ray spectrometer was added to the payload providing additional science return. This new instrument measured the solar 0.5--5 keV spectrum with 0.15 keV (FWHM) resolution \citep{2015ApJ...802L...2C}. Observations by this instrument fill an important gap in solar measurements in the $\sim$0.2--4 keV range. Recent analysis of the observations are suggestive of a ubiquitous high-temperature coronal plasma (5--10 MK) consistent the results obtained by EUNIS (\S\ref{sec-eunis}).

%% file: sec-astro-payloads.tex
\subsubsection{Diffuse X-ray emission from the Local galaxy (DXL) and X-ray Quantum Calorimeter (XQC)}
DXL \citep{dxl} 
and XQC \citep{xqc} 
both study the diffuse soft X-ray background. The soft X-ray background at 0.1--0.3 keV is dominated by emission from the ``local hot bubble'' (LHB) of $\sim$1.0$\times$10$^6$~K gas, a feature of $\sim$100 pc scale. Charge exchange on ions in the solar wind contaminate the LHB signal. At higher energies (0.3--1.0 keV), photons from (2--3)$\times10^6$~K gas distributed widely throughout the Galaxy dominate the spectrum. Understanding the spatial, temporal, and spectroscopic signatures of these components requires observing the soft X-ray background, with a high sensitivity over very large angular scales. The small fields of view of existing orbital observatories limits their ability to make these observations.
 

DXL has a large field of view and thin-window proportional counters with substantial effective area, combining to give it a grasp of 10 cm$^2$ sr at both 0.25 and 0.75 keV, several orders of magnitude higher than that of orbiting X-ray observatories.  It flew successfully in 2012, and an upgraded version is scheduled to fly again in December 2015.  The XQC experiment has a smaller grasp than DXL, but it is testing X-ray microcalorimeters. These new detectors have an energy resolution of $\sim$100$\times$ superior to that of DXL's proportional counters.  XQC has flown six times, increasing the heritage of X-ray calorimeters and helping pave the way for their use in orbital missions.

The few minutes of data from the XQC flights are currently the only astrophysical X-ray data obtained with microcalorimeters, but very similar detectors are installed on the Japanese observatory Astro-H, with a large telescope (but again a small field of view, designed for point sources) for studying black holes and clusters of galaxies at high spectral resolution. Astro-H is scheduled for launch in 2016.



\subsubsection{Far-ultraviolet Off Rowland-circle Telescope for Imaging and Spectroscopy (FORTIS)}

The FORTIS 
payload \citep{fortis} was recently deployed as a target-of-opportunity mission to observe Comet ISON as it neared the Sun.  FORTIS is an imaging far-UV spectrograph, and the goals of the ISON campaign were to measure the production of various volatiles (e.g. H$_2$O, CO, H, S) as the comet neared perihelion.  FORTIS observed ISON within a few degrees of the solar limb, much closer to perihelion than orbiting observatories were able to achieve, as they typically have sun avoidance zones within $\sim$45$^\circ$ of the Sun. Such a rapid response for a far-UV instrument with novel capabilities is only possible from a sounding rocket platform.

\subsubsection{PICTURE-B/C (Planetary Imaging Concept Testbed Using a Recoverable Experiment – Base/Coronagraph (PICTURE B/C)}
By testing new coronagraph technologies in a space environment, PICTURE-B/C \citep{pictureb} 
is a pathfinder for future missions which will directly image Earth-like planets around nearby stars (see article in this issue).  The first flight of PICTURE in late 2011 verified the ability of the system to maintain image stability at the required milliarcsecond level. The second launch late in 2015 should provide images of the planet-forming disk around the nearby star $\epsilon$-Eri.
The PICTURE payload is discussed further in this issue.

\subsubsection{Cosmic Infrared Background ExpeRiment (CIBER)}
CIBER\citep{ciber} measures the cosmic infrared background light in search of signatures from the first massive stars, whose UV light is redshifted down to the infrared. It has flown four times, with each successive flight honing the experiment as instrument and target characteristics became better known. The most recent flight was unusual in that it needed $\sim$2$\times$ longer exposure time than typically obtained, so it required a launch from WFF in order to use the larger Black Brant \rom{12} rocket instead of the more common Black Brant \rom{9} launched from WSMR. 

%% file: sec-future.tex

NASA's sounding rocket program is continuously evolving, as both new technology and new science ideas arise.  In particular, suggestions for new technology and innovation from the user community are discussed at bi-annual meetings of NASA's Sounding Rocket Working Group with the SRPO at Wallops.  

In the realm of astrophysics and solar rockets, the highest priority new technology initiatives that are being studied at this time include:

\begin{itemize}
\item Water recovery of telescope payloads which will enable astrophysics and solar payload launches from ranges other than WSMR.  This would enable higher apogees for such payloads (e.g., 500-600 km) using existing rocket motors, and hence considerably lengthen the available observation time, since range restrictions at WSMR typical limit apogees to about 400 km.  Water recovery for telescope payloads also opens up launch possibilities at new southern hemisphere locations as well as launches from WFF.
\item High altitude rockets (3000-3500 km) with larger diameters (e.g., 1-1.2 m) are also being considered.  Such vehicles would enable even longer observing times (e.g., 40 min) with larger aperture, more sensitive telescopes.
\item Very high telemetry capabilities, such as X-band telemetry that would enable rates of several hundred Mbps, as well as multi-TB storage of data on board for retrieval after flight.
\end{itemize}

The innovation, flexibility, reliability, quick turn-around, and low-cost of sounding rockets combine to provide an attractive platform for carrying out novel science in the space environment across multiple disciplines.  NASA's sounding rocket program has continued to enable cutting edge science in many fields of science and engineering, including astrophysics, solar physics, and geospace.  Since its inception, the program has continued to be vibrant and remains enormously popular in all disciplines. This is evidenced by the significant number of new science achievements and discoveries afforded by the program as well as the new science instrument technologies that have been tested on sounding rockets.  
